\documentclass[aps,prl,twocolumn,showpacs,amsmath,floatfix,citeautoscript]{revtex4}
\usepackage{graphicx}
\usepackage{float}
\usepackage{color}

\newcommand{\ket}[1]{| #1 \rangle}

\newcommand{\tx}[1]{\text{#1}}
\begin{document}

\hyphenpenalty=5000
\tolerance=1000

\title{Size consistency of tensor network methods for quantum many-body systems}

\author{Zhen Wang}
\affiliation{Key Laboratory of Quantum Information,
University of Science and Technology of China, Hefei, Anhui, 230026, People's
Republic of China}

\author{Yongjian Han}
\email{smhan@ustc.edu.cn}
\affiliation{Key Laboratory of Quantum Information,
University of Science and Technology of China, Hefei, Anhui, 230026, People's
Republic of China}

\author{Guang-Can Guo}
\affiliation{Key Laboratory of Quantum Information,
University of Science and Technology of China, Hefei, Anhui, 230026, People's
Republic of China}

\author{Lixin He}
\email{helx@ustc.edu.cn}
\affiliation{Key Laboratory of Quantum Information,
University of Science and Technology of China, Hefei, Anhui, 230026,  People's
Republic of China}

\date{\today }
\pacs{71.10.-w, 75.10.Jm, 03.67.-a, 02.70.-c}

\begin{abstract}

Recently developed tensor network methods demonstrate great potential for addressing the
quantum many-body problem, by constructing variational spaces with
polynomially, instead of exponentially, scaled parameters. Constructing such an efficient
tensor network, and thus the variational space, is a subtle problem and the main
obstacle of the method. We demonstrate the necessity of size consistency in
tensor network methods for their success in addressing the quantum many-body problem.
We further demonstrate that size consistency is independent of the entanglement
criterion, thus providing a general and tight constraint to construct the tensor network method.

\end{abstract}

\maketitle



%

The quantum many-body problem is one of the most fascinating topics in modern physics,
and as well as one of the most challenging \cite{anderson72}.
The challenges stem from the Hilbert space growing exponentially with the magnitude of the systems if they are
treated exactly. Fortunately, we do not need to address the entire Hilbert
space, because the physical properties of the quantum many-body systems are
determined by the ground state and some low excitation levels \cite{sachdev11}.
Therefore, it is possible to reduce the complexity of the problem
using a variational method, which considers the concerned states in a fixed space with
polynomially scaled parameters. However, constructing an efficient variational space for a many-body system,
which should include the states with which we are concerned, 
is a subtle problem and the main obstacle of the method. The recently developed
variational methods based on tensor network states (TNS), including the
matrix product states (MPS) \cite{verstraete08, ostlund95, klumper93,
  fannes92},
 and the projected entangled pair
states (PEPS) \cite{verstraete04, verstraete04_2}, offer a promising
scheme to construct such a variational space.
The success of the TNS relies on its satisfaction
of certain exact constraints.

One of the widely used constraints in TNS is the entanglement rule.
Entanglement is described by the block
entropy whose upper bound can be easily determined in TNS\cite{amico08}. In many of the 2D systems, the
block entropy of the ground state is assumed to satisfy the ``area law'', which
requires the block entropy of a region to be proportional to the area of its
boundary, not its volume. Therefore, the state in the variational space must satisfy the
same constraint. Under this rule, we can explain the success of the
MPS \cite{verstraete08, ostlund95, klumper93, fannes92}
(which is the variational space for the density-matrix renormalization group
(DMRG) \cite{white92, white93} method) for 1D many-body
systems and the failure of the MPS for the 2D
systems. PEPS \cite{verstraete04, verstraete04_2}, which is a natural development
of the MPS to 2D, can properly describe the ``area law'' for 2D systems.

However, the concrete entanglement character of a system is system dependent
and cannot always be known a priori. For some complicated systems, such as fermionic
systems, the block entropy of the ground state is beyond the area law, and
thus, the PEPS is no longer a proper variational space to approximate the ground
state \cite{wolf06, kraus10, verstraete06}.

In this Letter, we propose another
rule: the size consistency, which is an exact constraint to construct the
TNS. The size consistency is general and independent of the
Hamiltonian and characterizes the property of the variational space itself. 
We demonstrate that size consistency is essential for the success of the tensor network method and adds
more solid ground to the theory. The MPS/DMRG method is size-consistent in
1D but fails in 2D because of its lack of size consistency. However,
PEPS is size-consistent in any dimension. More importantly, we demonstrate
that a TNS satisfying the area law is not necessarily size-consistent. The
size consistency therefore provides an important and independent constraint
for the structure of tensor networks.

Size consistency is an important criterion that has been widely used in
quantum chemistry and condensed matter calculations
\cite{lowdin56,pople87,szabo_book}.
A size-consistent theory simply requires that the total energy of two non-interacting systems, $A$ and $B$, that is
calculated directly as a super-system, $A \oplus B$, should
be the sum of the energies of the two sub-systems calculated separately,
i.e., $E_{A \oplus B}=E_A+E_B$. We can define the size consistency error (SCE)
as $\Delta E= E_{A \oplus B}- (E_A+E_B)$ to evaluate the size-consistent
character. For a size-consistent theory, the SCE should be zero. The size
consistency condition is an exact and
general condition that a theory should satisfy, which is independent of the Hamiltonian and the
dimension of the system (and only determined by the structure of the
variational space in calculation). If a theory is not size-consistent, the quality of the
calculations decreases with increasing size of the system and eventually
breaks down. In other words, the size of the
variational space will dramatically (mostly exponentially)
increase to maintain the same precision with the increasing size of the system. In quantum
chemistry, the commonly used Hartree-Fock, and full configuration interaction
(CI) methods are size-consistent, whereas the
truncated configuration interaction method is notorious for its lack of
size consistency \cite{szabo_book}, and great efforts have been made to
overcome this problem \cite{pople87}.
Because size consistency is not bonded with the Hamiltonian or ground state,
this criterion provides an easy and general approach to verify the validity of
the methods.

\begin{figure}
\centering
\includegraphics[width=3.0in]{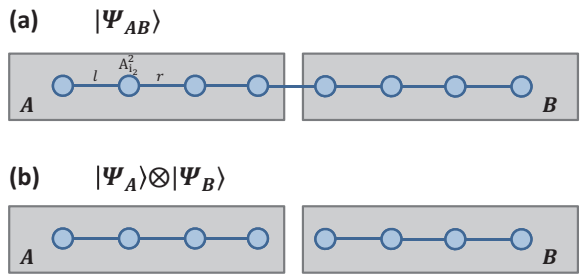}
\caption{(Color online) $A$ and $B$ are two independent 1D Heisenberg spin chains, each
  containing $L$-sites. (a) A schematic MPS representation of the $A \oplus B$
  super-system. (b) Schematic MPS representations of the sub-systems $A$ and $B$.
  Each filled circle represents a tensor $A^i_{i_k}$ and the line connecting them
  represents their virtual bonds $l$ and $r$.}
\label{fig:1dmpscut}
\end{figure}

We start to investigating the size consistency problem using the example of the 1D Heisenberg model (although the
size consistency problem is only dependent on the structure of the TNS itself
and independent of the Hamiltonian), $H=\sum_{\langle i,j \rangle} J_{ij}\,
S_i\cdot S_j$, where ${\langle i,j\rangle}$ represent the nearest
spin pairs, and $J_{i,j}$=-1 is the exchange interaction.
We study a spin chain of 2$L$ sites, illustrated in Fig.~\ref{fig:1dmpscut}.
The interaction between the $L$-th site and the $L$+1-th site is set to zero;
therefore, the system contains two non-interacting sub-systems, $A$ and $B$.
The ground state $\ket{\Psi}$ of the super-system
is expressed in MPS form \cite{verstraete08} [see Fig.~\ref{fig:1dmpscut}(a)],
\begin{equation}
\ket{\Psi_{\text{MPS}}} = \sum^{d}_{{i_1, \ldots
  i_{2L}}=1}\left(A^{i_1}(1)
\ldots A^{i_{2L}}(2L)\right)\ket{i_1, i_2, \ldots i_{2L}} \, ,
\end{equation}
where $d$=2 is the dimension of the physical indices ${i_k}$, and $A^{i_k}(k)$ are $D$ $\times$ $D$ matrices, where $D$ is the Schmidt cut-off,
except that $A^{i_1}(1)$ and $A^{i_{2L}}(2L)$ are matrices with dimensions $1\times D$ and $D\times 1$ at the boundary.
We obtain the ground state energy $E_{A \oplus B}$ of the super-system
using a variational MPS method\cite{verstraete08}.

Alternatively, we can calculate the ground state energies $E_A$ and $E_B$
of the sub-systems $A$ and $B$ separately [see Fig.~\ref{fig:1dmpscut}(b)]
and add them to get the total energy of the 2$L$ spin chain.
We calculate the SCE for 2$L$ up to 128 sites, using the Schmidt cut-off $D$=1
- 24. We observe that the SCE $\Delta E$=0 for all the spin chain lengths and
the Schmidt cut-off $D$, demonstrating that MPS is indeed size-consistent in
1D. Because it has been demonstrated that the DMRG method is equivalent to the
variational MPS method, the DMRG method is also size-consistent in 1D.
This implies that the MPS/DMRG method can maintain the accuracy of the
calculations for larger systems without (at least significantly) increasing
$D$, explaining its success in 1D.
This result is in contrast with the numerical renormalization group method
 \cite{wilson75}, in which one continuously truncates the global
variational spaces, and therefore, the variational spaces of the super-system
and sub-systems are different, leading to the size inconsistency and failure
of the method.

\begin{figure}
\centering
\includegraphics[width=3.0in]{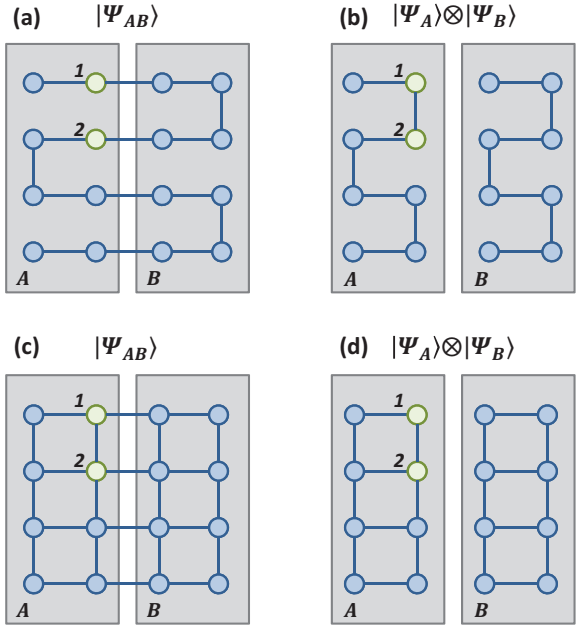}
\caption{(Color online) $A$ and $B$ are two independent 2D Heisenberg spin lattice, each
  containing $L \times 2L$-sites.
  (a) A schematic MPS representation of the $A \oplus B$
  super-system.
  (b) Schematic MPS representations of the sub-systems $A$ and $B$.
  (c) A schematic PEPS representation of the $A \oplus B$ super-system.
  (d) Schematic PEPS representations of the sub-systems $A$ and $B$.
}
\label{fig:2dcut}
\end{figure}

We now move to the 2D case.
We study the 2$L \times$2$L$ Heisenberg spin lattice presented in
Fig.~\ref{fig:2dcut}, which contains two non-interacting $L \times 2L$ lattices denoted by $A$ and $B$,
as the exchange interactions $J$ are set to zero between the
two systems as performed for the 1D spin chain.
We compare two kinds of tensor networks: one is a
direct generation of MPS, and the other is PEPS.

\begin{figure}
\centering
\includegraphics[width=2.8in]{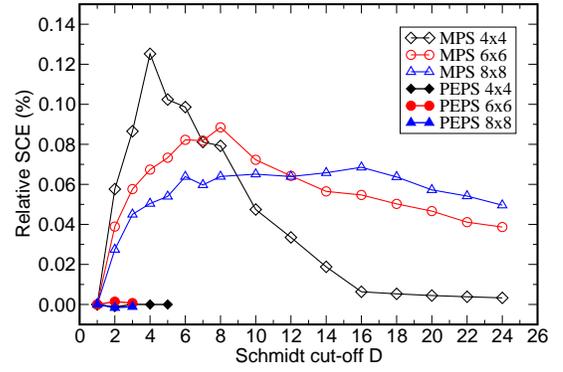}
\caption{(Color online) The relative SCEs of the 2D Heisenberg model as functions of the Schmidt cut-off for different lattice sizes.
The relative SCEs of MPS are represented by in open labels, whereas the relative SCEs of PEPS are represented by solid labels.}
\label{fig:2dsc}
\end{figure}

We first calculate the ground state energies of the 2$L$$\times$2$L$
lattice and $L$$\times$2$L$ lattice and their SCEs using the variational MPS method.
When calculating the ground state of the $A \oplus B$ super-system,
the MPS is arranged in a snaky manner as illustrated in
Fig.~\ref{fig:2dcut}(a), 
whereas when we calculate the ground states of $A$ and $B$ separately, the MPS
is arranged in the manner illustrated in Fig.~\ref{fig:2dcut}(b).
The relative SCE, $\Delta E/E_{A \oplus B}$, for 2$L$=4, 6, 8 and Schmidt
cut-off $D$=1 - 24 are shown in Fig.~\ref{fig:2dsc}. For all the lattice
sizes, the SCE $\Delta E$=0 at $D$=1.
This result occurs because at $D$=1, the MPS is a direct product state,
resembling the Hartree-Fock method, which is size-consistent.
However, for $D>$2, the SCE first increases with $D$ up
to a certain value $D_c$, and then decreases as $D$ increases.
As shown in Fig.~\ref{fig:2dsc}, $D_c$ increases with the system size,
and more importantly, the SCE decays much
slower as the system size increases. For the 4$\times$4 system, the SCE becomes
small for $D >$16, whereas for the 6$\times$6 and 8$\times$8 system, the
SCEs are still significant even for $D$=24.
These results suggest that to maintain the accuracy of the method, $D$ must increase rapidly with the system size.
Indeed, it has been observed that to achieve certain accuracy,
the Schmidt cut-off $D$ must grow exponentially with the dimension of the
lattice \cite{liang94}, leading to the failure of the MPS/DMRG methods at in 2D.

PEPS is a natural generalization of MPS to two
and higher dimensions from entanglement insight, whose entanglement is bonded
by the number of entangled pairs and is determined by some projectors (denoted
by some tensors) \cite{verstraete04}.
A PEPS variational wavefunction for the $2L\times 2L$ system can be
written as
\begin{equation}
\ket{\Psi_{\tx{PEPS}}} = \sum^d_{i_{1,1},\ldots,i_{2\tx{L},2\tx{L}}=1}
C\left(\left\{A^{i_{m,n}}(m,n)\right\}\right) \ket{i_{1,1},\ldots,i_{2\tx{L},2\tx{L}}}.
\end{equation}
Similar to MPS, $d$ is the dimension of the physical indices ${i_{m,n}}$. The
tensors $\{A^{i_{m,n}}(m,n)\}$ are now four index tensors $A_{u,d,l,r}$ that
connects to their neighbor tensors in the up, down, left and right directions
with bond dimension $D$, except for the tensors at the four borders of the
lattice. The function
$C$ contracts all the virtual indices $u$,$d$,$l$,$r$ of all the tensors.
$D$ is the dimension of the virtual states. We obtain the ground state using a
variational method following Refs. \cite{isacsson06}.

PEPS has been demonstrated to be successful in treating 2D many-body
systems \cite{verstraete04, murg07, isacsson06, gonzales12, ji11, corboz11}.
For the Heisenberg model studied here,
the PEPS method converges very well to the exact ground state
even with a small Schmidt cut-off $D$=4, for the 4$\times$4 lattice.
The relative SCEs, $\Delta E/E_{A \oplus B}$,
of the PEPS are also presented in Fig.~\ref{fig:2dsc} and
compared with those of the MPS method for 2$L$=4, 6, 8.
Unlike the MPS, we observe that the SCEs of PEPS are all zero within numerical
error, regardless of the size and Schmidt cut-off $D$, confirming that PEPS is
indeed size-consistent. This implies that we can use a relatively small
Schmidt cut-off even for a large system without losing accuracy.

With the former intuitive results, we can now discuss the general size consistency of the TNS.
To be a size-consistent tensor network theory, the direct product of
any wave functions $|\Psi_A \rangle$ and $|\Psi_B\rangle$ (for two
non-interacting sub-systems $A$ and $B$,respectively), which are represented in the tensor network of some fixed bond dimension $D$,
should be presented exactly by a tensor network of the super-system $\Psi_{A \oplus B}$ with the same $D$.
For the MPS case, this can be done in 1D systems but not in
2D systems. More explicitly, the states of the sub-systems $A$ and
$B$, $|\Psi_A \rangle$ and $|\Psi_B\rangle$, can be represented in the form of
Eq.(1) with tensors $A^{i_1}(1), \cdots, A^{i_L}(L)$ and $B^{i_1}(1), \cdots,
B^{i_L}(L)$, respectively. Note that the direct product of the two
sub-systems can be represented in MPS form with $D$=1. Therefore, we can construct a MPS for the supersystem in the form of Eq.(1)
with tensors $C^{i_1}(1),\cdots,C^{i_{2L}}(2L)$ (with bond
dimension $D$), which are defined as:
\begin{eqnarray}
\label{eq:directp}
C^{i_k}(k) &=& A^{i_k}(k),  \, \text{when}\, k=1,2,\cdots,L-1 \\ \nonumber
C^{i_L}_{l,1}(L) &=& A^{i_L}(L), \, \text{else}\, C^{i_L}(L)==0 \\ \nonumber
C^{i_{L+1}}_{1,m}(L+1) &=& B^{i_L}(L), \,\text{else}\, C^{i_{L+1}}(L+1)==0 \\ \nonumber
C^{i_{L+k}}(L+k) &=& B^{i_L}(k), \,\text{when}\,  k=1,2,\cdots,L-1
\end{eqnarray}
This MPS is in the super-system $\Psi_{A \oplus B}$ with the same dimension
$D$ and exactly equal to the direct product of $|\Psi_A \rangle$ and
$|\Psi_B\rangle$.

A similar construction can be used in two dimensions; however, the resulting
state is not in the MPS space in Fig. 2 (a) and (b), because there are some
additional bonds in the new state beyond MPS, such as the bond between sites 1
and 2. Therefore, there are two MPS states in the subsystem, whose direct
product is not in the MPS space of the super-system, and the MPS method is not
size-consistent. The physical reason why MPS is not size-consistent in 2D
is described as follows. The entanglement between sites 1 and 2 in sub-system
A has to go through B, i.e., 
the entanglement in a system un-physically depends on the states of another
system that has no interaction with the system at all. 
In this case, if $\Psi_{A\oplus B}$ is written as the direct product of $\Psi_A$ and $\Psi_B$,
similar to Eq.~\ref{eq:directp},
there would be no entanglement between sites 1 and 2 at all.

Fortunately, this result would not occur for the PEPS in 2D.
By performing the same trick as in the 1D MPS for all edges linking the two
sub-systems in Fig.~\ref{fig:2dcut}(c) and using
the same tensors at all other sites as those of the sub-systems. The resulting
state is exactly equal to the product state of the subsystems and in the PEPS
space with the same bond dimension $D$.

From the discussion above, we find that the size consistency of a tensor
network method can be reduced to the geometry structure of the network and is
independent of the system Hamiltonian. The network should have the same
structure when two sub-lattices merge into a super-lattice or a super-lattice
separates into two sub-lattices, i.e., no additional tensor bonds beyond those
in the super-lattice should be presented in the sub-lattice networks.
This simple criterion can readily exclude many of the tensor
networks. We also note that whereas the truncated CI method
suffers from its lack of size consistency, the great advantage of tensor
network methods is that much easier to construct a size-consistent theory based on
TNS.

The entanglement rule and the size consistency rule originates from the
different points of view. In fact, the entanglement is concerned with
the global property of the state, whereas the size consistency criterion,
defined by the energy relation, is simply determined by the local reduced
density matrices.
The relation between the global state and its local reduced density matrices
is very complicated and subtle\cite{coleman63}. A theory satisfying the
size consistency condition may have no
entanglement, e.g., the Hartree-Fock and Gutzwiller methods
\cite{gutzwiller63}, whose variational spaces are product states.

\begin{figure}
\centering
\includegraphics[width=3.2in]{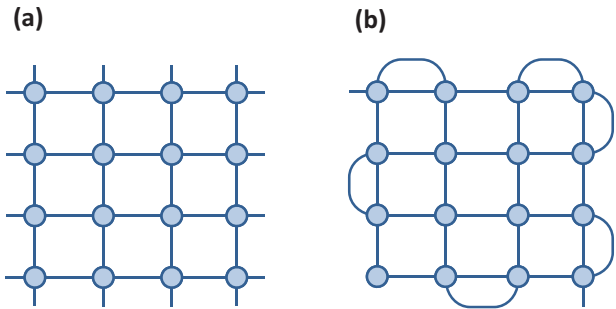}
\caption{(Color online) The two types of string bond states (SBSs)
  proposed in Ref.\cite{schuch08}:
(a) horizontal and vertical lines, and (b) a long line covers all the sites.}
\label{fig:sbs}
\end{figure}

On the other hand, a theory whose states in the variational space satisfy the
area law is not necessarily size-consistent. 
Figure~\ref{fig:sbs} depicts two types of string bond states (SBSs) proposed by Schuch et al. \cite{schuch08},
which both satisfy the area law.
Whereas the lines SBS in Fig.~\ref{fig:sbs}(a) is size-consistent,
the long line SBS in Fig.~\ref{fig:sbs} (b) is not. This
can be seen by dividing the lattice in Fig.~\ref{fig:sbs}(b) into two
parts, the long line SBS of the sub-systems have additional bonds beyond those in the super-system.

We note that size consistency is only a necessary but not a sufficient
requirement for a ``good'' theory. Size consistency describes only
the additivity property of the energy. If we are
only interested in the energies, even a low level
a size-consistent theory, such as the Hartree-Fock method or Gutzwiller method
etc., may provide a very good approximation to separate the quantum phases
from energy.
However, in many cases, we are often interested in more subtle physical
properties such as the correlations of the ground states, for which
these theories may fail. Therefore,
a good tensor network method for quantum many-body problems should
satisfy the following constraints: (i) The
theory must be size-consistent; (ii) The theory must satisfy proper
entanglement scaling with system sizes; (iii) The theory must be systematically
improvable via some controlling parameters and eventually converge to the
exact results; and finally (iv) The theory must be efficient, such that the
computational cost scaling
with the controlling parameters is modest.

To summarize, we have discussed the size consistency of tensor network
methods for quantum many-body systems.
We demonstrate that the size consistency is essential for the success
of the tensor network method for the quantum many-body problem, which is
responsible for the failure of MPS and the success of PEPS in 2D.
We further demonstrate that a tensor network state satisfying the area law is
not necessarily size-consistent. Size consistency
therefore provides an independent and tight constraint in constructing the
tensor network method. 
We propose four criteria for a
good tensor network method for the quantum many-body problem.

LH acknowledges the support from the Chinese National
Fundamental Research Program 2011CB921200, and the
National Natural Science Funds for Distinguished Young Scholars.
YH acknowledges the support from the Central Universities WK2470000004, 
WK2470000006, WJ2470000007 and NSFC11105135.

\bibliography{refs}

\end{document}